# Consistency Check of Segregation Energies of Elements at Ni and Fe Grain-Boundaries

*Author:* Wilfried Wunderlich 1),
1) Tokai University, Faculty of Engineering, Department for Material Science
259-1292 Hiratsuka, Kitakaname 4-1-1, Kanagawa-ken, Japan
E-mail address: wi-wunder@rocketmail.com

**Abstract.** The need for advanced functional materials is expected to provide a boost in powder metallurgy, where the impurities on powder surfaces is incorporated as at grain boundary segregation. This paper has three aims. First, we analyze whether the reported data of Ni and Fe hosts can be correlated to basic thermodynamic data on chemical elements. The second aim is the consistency check, which is suggested to be applied for any data base. The third aim, is whether information of 50 most important elements can give additional information for prediction of unknown data. The data are analyzed whether a set of basic data is sufficient for characterizing the segregation. The data of the solvents were analyzed using the software R for principal component analysis (PCA). We grouped and correlated the data to Mendeleev number and thermodynamic data on pure elements. As a result, we found that the embrittlement depends strongly on the chemical bonding, but weakly on mechanical factors. Surprisingly, the geometry of the grain boundary type such as interlayer distances, and local atomic volumes has only a minor influence.

Introduction

Grain boundary (GB) segregation in Ni and Fe alloys has become main research topic for many experimental studies as well as DFT-calculations and the data are summarized in overview papers [1-5] as well as the Segrocalc database [6]. Segregation influences materials properties remarkably, most investigated is embrittlement [1-4, 7]. This research will gain more interest in future, as free surface (FS) segregation on powder particles will contribute intrinsically to the GB segregation during additive manufacturing (AM) [8,9]. While data on more and more material combinations became available over the last decade, progress was also made on characterization of segregation. It has been well established that segregation is characterized by three independent parameters [1-4], the GB segregation energy $E\_GB$, which is the chemical affinity for the solute atom to accumulate in the grain boundary region rather than in the crystal. DFT calculations require seven different steps to acquire all data [2]. Some systems form interstitials, while at most atom combinations the foreign atoms sits as substituents on crystal lattice sites of the host. Furthermore, we need the surface segregation energy $E\_FS$ and the strength of embrittlement $E\_Steem$, counted positive for strengthening and negative for weakening the atomic bonds.

As summarized in an overview paper, these three quantities depend not only on the type of segregation atom and its host, also on the concentration and on each individual segregation site, which yields to a bell-shaped distribution of segregation sites recently named as density of segregation sites [5]. We restricted our search on data measured at room temperature, while being aware that temperature could favor one potential site over another one . Usually the Sigma-5 grain boundary is used for DFT model calculations as the GB which possess the most pronounced structural unit at the grain boundary. The recent progress in machine learning (ML) allowed researchers to calculate segregation phenomena and compare experimental data with different calculation methods. After having gained experience in correlating the electron-phonon interaction parameter (EPI) more systematically [10], we try the similar method in this paper. The key for this correlation is emphasized in the so-called Linus data file. The periodic table of elements can be ordered in two ways, either as the usual ordering by ascending atomic number Z equal to the number of all electrons, or the vertical ordering using the so-called Mendeleev number (MN) [11], which takes into account the fact that the chemical bonding strongly depends on the number of outer electrons.

In this paper we investigate the consistency of the three segregation energies E_FS, E_GB and E_Steem with other element specific properties. We describe the calculation method, the correlation with other data, the achievements and then analyze the data consistence. The results allow suggestions for data revision, while other data are suggested as prediction of values for yet unknown data, and need later to be confirmed by proper experiments or calculations. In the discussion we provide guidelines for future research.

**Calculation method**

The data analysis follows a procedure consisting of several steps similar as in machine learning. At first literature data for E_FS, E_GB and E_Steem for the host atoms Ni and Fe were collected from the overview papers of the main three groups and converted into the unit [eV/atom]. When necessary, we also examined the original papers. In some systems, when the different sites may yield to a distribution of energy values for each site, we have chosen the most likely average values mostly from Sigma-5 GB's. We are aware that a detailed ML approach of all available data could gain more potential insight in GB segregation phenomena, but the simplification allows a faster conclusion for an overview. The result of this consistency check confirmed excellent correlation between the different literature data which support each other.

The second step during this machine learning approach is the grouping of the data, for which we used the software "R" and "Weka" and their visualization tools. The preliminary test with the restricted set of around 15 labelled data available at that time [2-4] was published earlier [9] with a classification into six groups, named after their corresponding atom groups alkali, transition metals, noble metals, semiconductors, "Fluor" naming the halogen group 17 and "gas" naming the noble gas (group 18). While this approach worked well for evaluating the dependence of EPI, the present set of around 100 labelled data allows a much more precise correlation as reported and discussed in the following section. We had to expand the six groups and differentiate into groups of almost all 18 rows of the periodic table. The next step is fitting the data to a model. Surprisingly we found a linear dependence with MN, while other correlations are described in the next section. Also the comparison between data on Ni and Fe improved the data analysis. After finding these correlations, the last step is to predict data of yet unknown element combinations. We restricted the prediction to interpolation, as experience in ML shows that extrapolation is usually considered as not reliable enough.

**Results and discussion**

The segregation data strength of embrittlement E_Steem for Ni as host are displayed in fig. 1 on x-axis, while the Mendeleev number MN is drawn on y-axis, starting with 1 for Li, 2 for Na and so on. Excluding H, the Alkali elements marked with "1" in red for first row follow a linear behavior in the strength of embrittlement, where a positive value means strengthening the grain boundary. Second (Alkaline Earth) and third row elements follow the same slope of 0.32 eV/atom as well as constant intercept distance. Hence, we have fitted the segregation behavior of 13 elements. The E_Steem values for Fe as host are shown in fig. 2 and show almost identical behavior.

Returning to Fig. 1, the values for Lanthanides from La to Sm are displayed in dark brown. The negative slope is obviously influenced by their f-electrons. The slight change in slope between Sm and Eu is caused by the sudden change in EPI [10], and we mark it with a change into blue color.

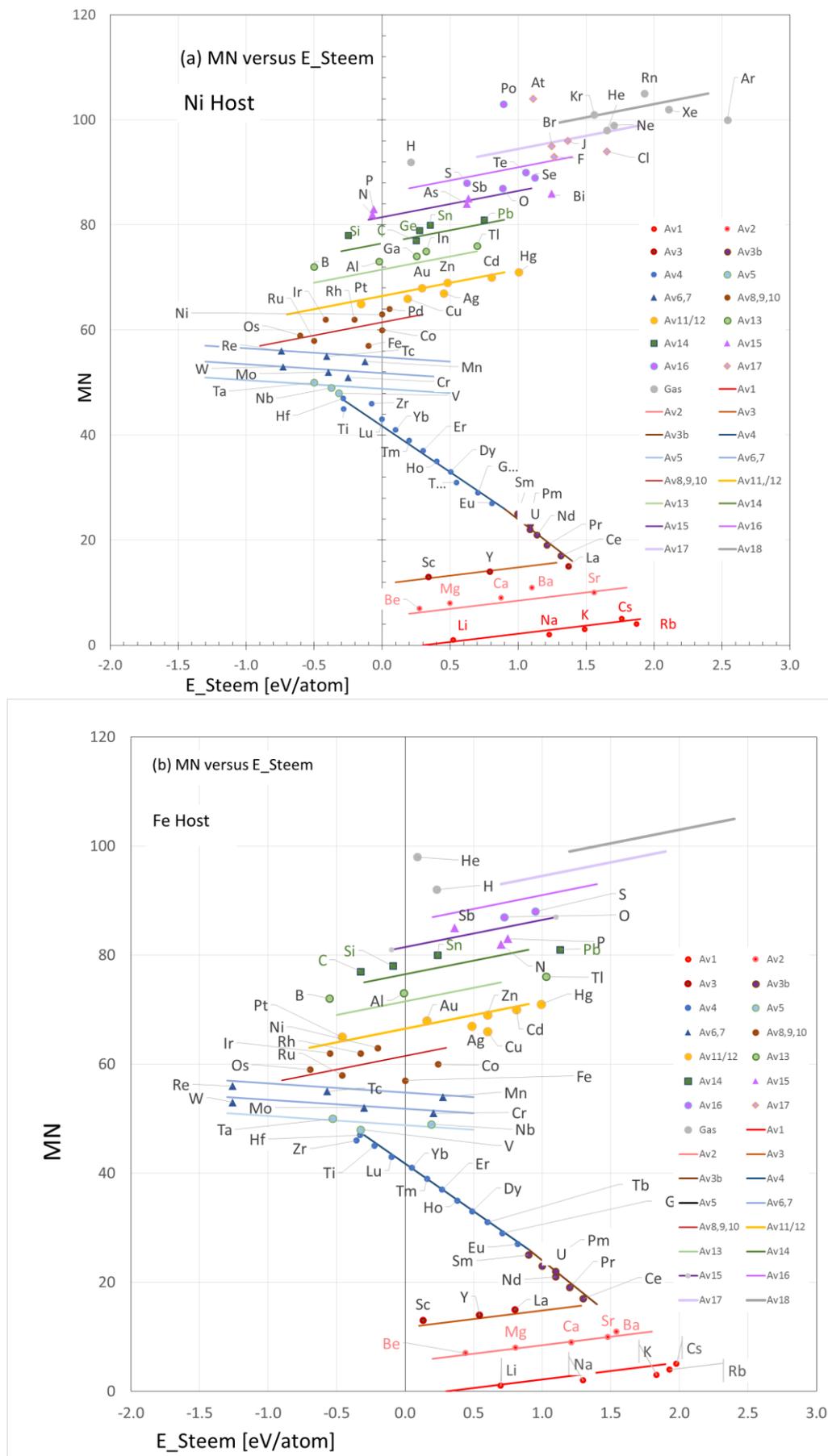

Fig. 1 Strength of embrittlement on x-axis versus the Mendeleev number MN for segregation of foreign atoms at (a) Ni, (b) Fe grain boundaries.

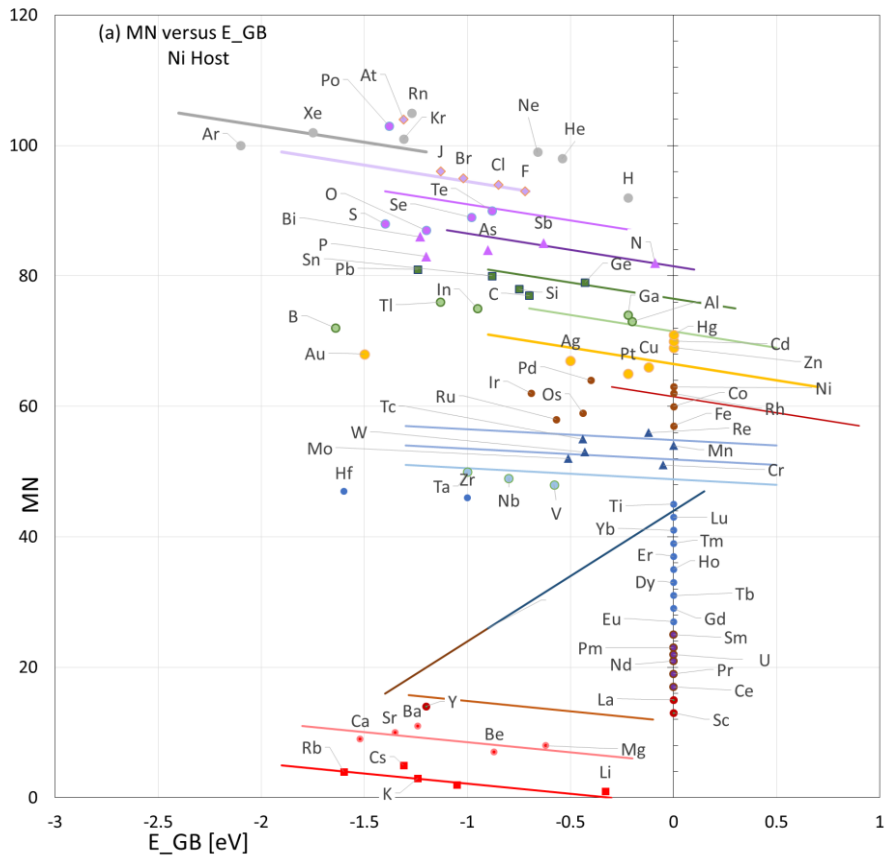

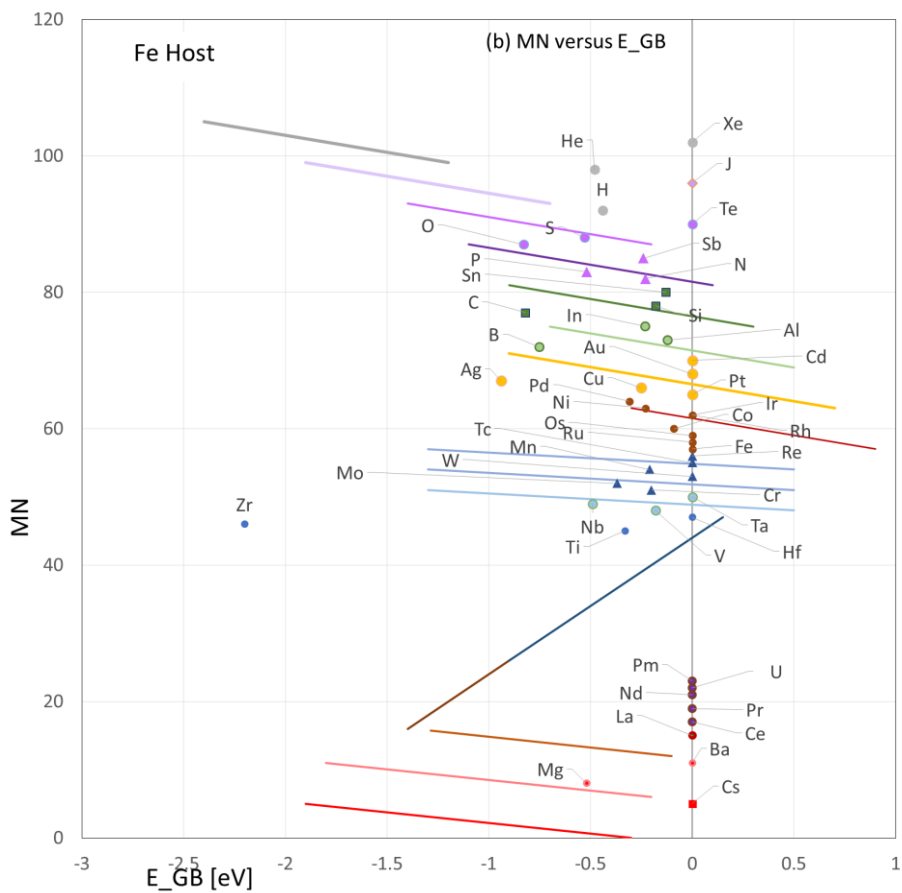

Figure 2 Segregation energies on x-axis versus the Mendeleev number MN for segregation of foreign atoms at (a) Ni, (b) Fe grain boundaries..

The MN is alternating between Lanthanides and Actinides, hence allowing to predict that the yet unknown E_Steem value for example for U should lie between that of Nd and Pm. The behavior of E_Steem for 4th row elements Ti, Zr, Hf is not so clear, but obviously follow that of Lanthanides, while the 5th (V,Nb,Ta) and 6th row (Cr,Mo,W) transition elements can again fitted with same slope and intercept as displayed in their light blue color. E_Steem values of elements lying in the 8th, 9th and 10th row are lying almost on the same line, while the noble elements (11th and 12th row) marked with orange color have significant larger E_Steem values. Elements on all other rows of the periodic table (13th to 18th) can be again fitted with same slope and intercept, although the experimental values show slight deviations is some cases, for example F, Cl and Ar. Elements which cannot be predicted with this MN-calibration procedure are H, Po, and At.

The situation for Fe as host as displayed in Fig 2 is very similar, as exactly the same slopes of E_Steem versus MN fits the available data well. In many cases such as the Lanthanides, the data are not yet available. In the case of Mo and W E_Steem for Ni as host is 0.9 eV, while they are 1.15 eV for Fe as host. Also the fitting for elements beyond noble metals is not as good as in the case of Ni as host. More experimental data are needed to confirm a most likely systematical shift, which we indicate with hatched lines. Nevertheless, the data fitting can predicts at least the range for E_Steem values for some not yet measured elements, which as summarized in a table at the end.

The success of the ML-based analysis allows us to proceed to E_GB as for Ni and Fe as host as shown in fig 3 using the same color code as in the figures before. While less data have been measured and displaying the yet unknown data with a value of zero, the fitting procedure is not so reliable, but shows almost the same behavior yet with opposite slope. For both hosts, Fe and Ni, we can again proclaim the same tendency. As indicated by the hatched lines, the present data would not allow a reliable extrapolation, but the overall "big picture" justifies to publish these "most likely" data, with the option that future measurements would hopefully confirm these data.

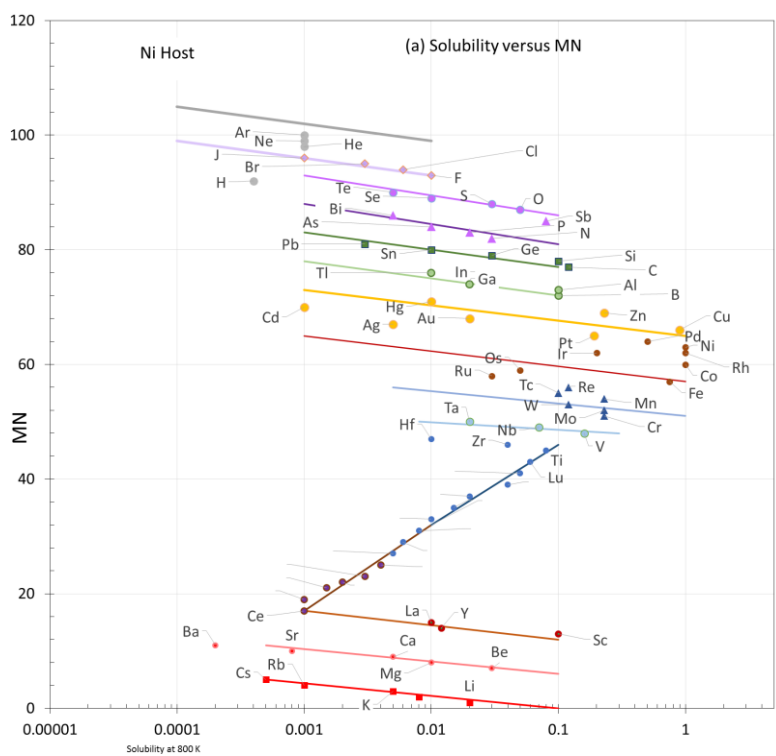

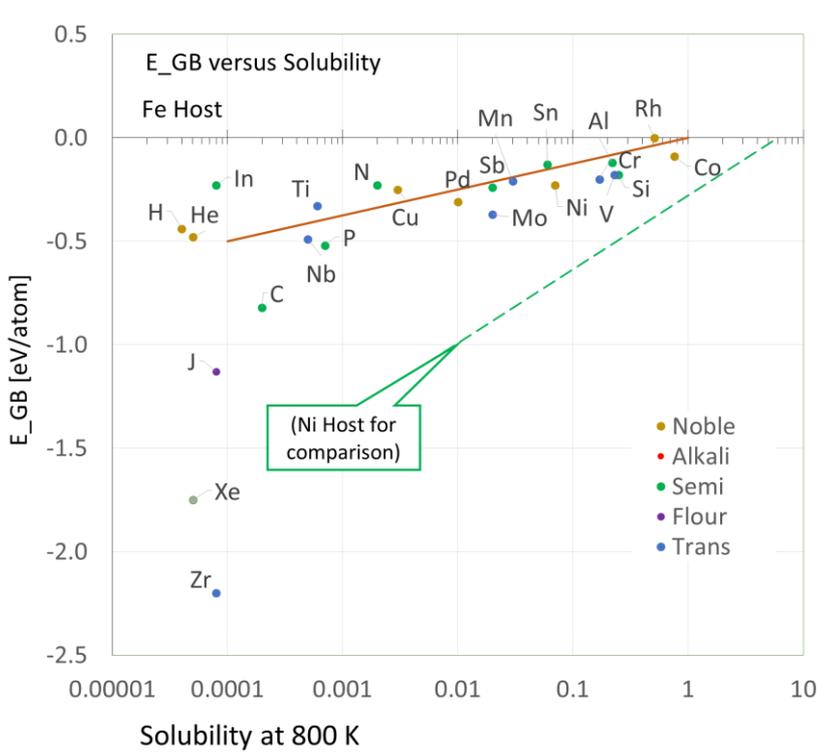

Figure 3 Logarithm of solubility at 800K of foreign atoms (a) in Ni versus the Mendeleev number MN, and (b) in Fe versus the segregation energy at grain boundaries E_GB.

A similar analysis was performed for E_FS, the third parameter for characterizing segregation, which quantifies the segregation energy of foreign elements at free surfaces. We have performed the same analysis and found a similar behavior of E_FS on the dependence of MN, except that the range of energies is much larger, ranging from -6.4 eV for Cl, Br and Ba and +1.0 eV for W in Ni or -2 eV for Sc, and +1.48 eV for Re in Fe. This figure is not shown due to space restrictions.

The strength of the described method is, that all the three segregation data E_Steem, E_GB and E_FS, for the foreign atoms in the same column of the periodic table depend linearly on MN with a shallow slope. In contrary, slight errors in the training data could spoil the reliability of the whole prediction. The model established in this paper is consistent for Ni and Fe, which are elements rather close to each other in the periodic table. Nevertheless, this finding should be double-checked by another parameter. In a previous paper [8] we suggested a correlation between EPI and E_Steem and the present analysis can confirm this. The data are not shown due to space limitation. This check suggests, that Ar is the element with the highest EPI value among all elements, yielding to EPI =2.8.

Figure 3 Logarithm of solubility at 800K of foreign atoms in Ni versus (a) the Mendeleev number MN, and (b) versus the segregation energy at grain boundaries E_GB.

Lejcek et al. suggest a correlation of E_GB with the solubility at 800 K [1], which can be estimated with Calphad calculations or from experimentally measured phase diagrams. We draw the provided data [1] for Ni against MN and E_GB as shown in fig. 3 (a) and (b), where solubility is shown in logarithmic scale. The negative slope in fig, 3 (a) confirms the intuitive assumption that atoms with smaller ionic radii have a higher solubility. The data in fig 3 (b) show a large scattering, but the correlation for higher solubility seems obvious. The slope for data on Fe, not shown here due to space limitation, shows a significant smaller slope as indicated by the hatched line.

With the knowledge gained by this analysis it is possible to apply an optimization procedure for all data based on minimizing the overall deviation of the least squares in order to predict reliable values, as soon as this method has been widely accepted. If there is an urgent need to get an estimation of a specific data point, the present figures might be a proper guideline. They can also be used for consistency check whether present measurements of E_Steem, E_GB or E_FS need to be revised. The presented results are expected to stimulate further research in the field of grain boundary segregation as well as materials development.

Summary

The quantitative estimation of segregation energies is an important field in ongoing research. We suggest a consistency check, whether the measured or DFT-calculated data are correct, by plotting the data versus the Mendeleev number MN as a reliable and powerful tool. The strength of embrittlement E_Steem follows the ordering in rows within the periodic table for both considered host elements Fe and Ni in almost the same behavior. The GB segregation energy E_GB can also be correlated with MN allowing to predict the values for yet unknown values such as for Lathanides. Furthermore, E_GB shows good correlation with the solubility of foreign atoms.